\documentclass[twocolumn,showpacs,prl,aps,amsmath,amssymb]{revtex4}

\usepackage{graphicx}
\usepackage{dcolumn}
\usepackage{bm}

\begin{document}
\title{ Anomalous electron-phonon interaction in doped LaOFeAs: a First Principles calculation }

\author{Felix Yndurain}
\email{felix.yndurain@uam.es.}
\affiliation{Departamento de F\'{i}sica de la Materia Condensada
and Instituto de Ciencia de Materiales "Nicol\'{a}s Cabrera".
Universidad Aut\'{o}noma de Madrid. Cantoblanco. 28049 Madrid.
Spain.}

\author{Jose M. Soler}
\affiliation{Departamento de F\'{i}sica de la Materia Condensada
and Instituto de Ciencia de Materiales "Nicol\'{a}s Cabrera".
Universidad Aut\'{o}noma de Madrid. Cantoblanco. 28049 Madrid.
Spain.}

\date{\today}

\begin{abstract}
We present first principles calculations of the atomic and  electronic structure of electron-doped LaOFeAs. We find that whereas the undoped compound has an antiferromagnetic arrangement of magnetic moments at the Fe atoms, the doped system becomes non magnetic at a critical electron concentration. We have studied the electron-phonon interaction in the doped paramagnetic phase. For the $A_{1g}$ phonon, the separation between the As and Fe planes induces a non-collinear arrangement of the Fe magnetic moments. This arrangement is anti parallel for interactions mediated by As, and perpendicular for Fe-Fe direct interactions, thus avoiding frustration. This coupling of magnetism with vibrations induces anharmonicities and an electron-phonon interaction much larger than in the pure paramagnetic case. We propose that such enhanced interactions play an essential role in superconducting compounds close to an atiferromagnetic phase transition.
\end{abstract}


\maketitle

Very recently, Kamihara and co-workers \cite{Kamihara-2,Kamihara-1} have reported a new family of Fe based compounds  that are superconducting when doped. All these ROFeAs (R=La, Sm, etc) materials are formed by FeAs layers separated by insulating rare-earth-oxide layers. They are superconducting when doped with electrons by substituting O for F, whereas the parent compounds are antiferromagnetic (for a recent review see \cite{Review}). The superconducting critical temperature $T_c$ is as high as 50 K \cite{50K}. The phase diagram has been well established experimentally both as a function of doping \cite{SmFeAsO-Phase-Diagram,LaOFeAs-Phase-Diagram} and pressure \cite{Pressure-Phase-Diagram}. The variation and ordering of the antiferromagnetic phase with temperature has been measured with neutron diffraction \cite{clarina-nature}. The phonon modes have been determined by Raman spectroscopy \cite{Raman}. Photoemission experiments, and first principles calculations, have revealed a Fermi surface with two electron pockets and two hole pockets \cite {Scalapino}. 

\begin{figure}[ht]
\includegraphics[width=66mm]{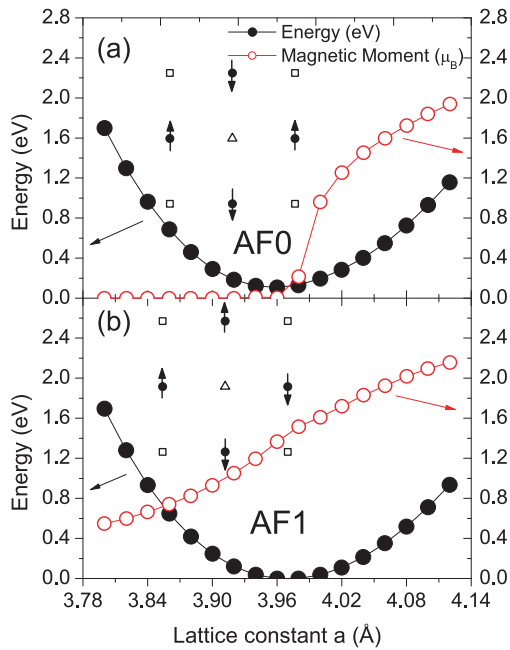}
\caption{(Color online) Variation of total energy (left scale) and magnetic moment of Fe atoms (right scale) with the lattice constant $a$ in two different antiferromagnetic arrangements of LaOFeAs: (a) direct Fe-Fe antiferromagnetic order (AF0); and (b) As-mediated antiferromagnetic order (AF1). The zero of energy is at the AF1 minimum. The orientations of the Fe magnetic moments in each configuration are indicated in the insets, where squares and triangles stand for As atoms above and below the plane of Fe atoms. The total energy is for a $(\sqrt{2}a\times\sqrt{2}a \times{c})$ supercell.} \label{figure1}
\end{figure}

The precise mechanism responsible for the superconductivity has not been yet established. Theoretical calculations \cite {Boeri} rule out electron-phonon interaction to be the only pairing mechanism. Electron-phonon interaction in the paramagnetic phase can only account for a maximum $T_c$ of 0.8 K \cite {Boeri}. Other calculations \cite{haule} also rule out phonon mediated superconductivity. The magnetic structure of the parent compounds is well established experimentally although, from the theoretical point of view, there are some discrepancies, between different calculations and with the experimental results, in the magnitude of the magnetic moment at the Fe atoms \cite{Korotin}.

In this work we present first principles calculations to analyze in more detail the electron-phonon interaction in the doped paramagnetic phase. The Density Functional \cite {DFT1, DFT2} calculations are performed using the SIESTA code \cite {Siesta1, Siesta2} which uses localized orbitals as basis functions \cite{Orbitals}. We use non-local norm conserving pseudopotentials and a Local Density Approximation (LDA) for the exchange and correlation functional. The calculations are performed with stringent criteria in the electronic structure convergence (down to $10^{-5}$ in the density matrix), Brillouin zone  sampling (up to 18000 $k$-points), real space grid (energy cut-off of 500 Ryd) and equilibrium geometry (residual forces lower than $10^{-2}$ eV/\AA). Due to the rapid variation of the density of states at the Fermi level, we used a special smearing method \cite{smearing}.

\begin{figure}[ht]
\includegraphics[width=68mm]{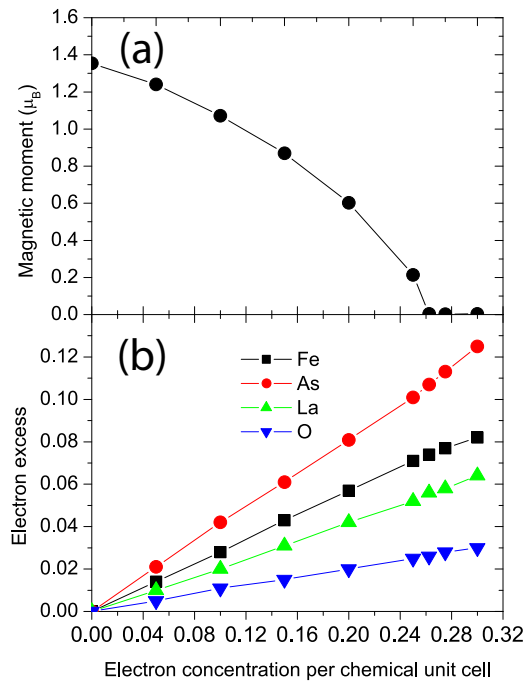}
\caption{(Color online) Calculated effects of electron doping in the AF1 magnetic configuration of LaOFeAs (see text). (a) Magnetic moment per Fe atom. (b) Distribution of the extra charge among the different atoms. } 
\label{Doping}
\end{figure}

We first calculate  the variation of the total energy with the lattice parameter for various possible magnetic configurations. In Figure \ref{figure1} we show the results for two possible antiferromagnetic configurations in the Fe atoms: direct antiferromagnetic ordering between nearest neighbor Fe atoms (AF0) and antiferromagnetic Fe-Fe interaction mediated by As (AF1). We find that, at the calculated equilibrium lattice constant (3.97\AA), the ground state has AF1 magnetic order. Its energy difference with AF0 (which is non magnetic at that lattice constant) is 26 meV per formula unit. Ferromagnetism and other non-collinear magnetic configurations are found to be unstable. The AF1 magnetic moments are $1.3 ~\mu_{B}$ per Fe atom, in reasonable agreement with  previous all-electron density functional calculations \cite {Yildirim-1,singh}. This calculated value is larger than the 0.36 $\mu_{B}$/atom found in neutron diffraction experiments \cite{clarina-nature}, possibly due
to neglected spin fluctuations \cite{Korotin}.

In order to simulate the doped phase we have added electrons to the system while keeping the charge neutrality by adding a uniform positive charge background. As shown in Figure \ref{Doping}, the magnetic moments decrease with electron doping, and they disappear above a critical concentration. This critical value (0.26 electrons per formula unit) is larger than the experimental one, probably due to the overestimation of the magnetic moment in the undoped phase. From the structural point of view, the main effect of electron doping is to decrease the distance between the As and Fe planes. For instance, in the paramagnetic phase, with 0.275 extra electrons, the compression is 0.02 \AA. This compression, added to the filling of the Fe $d$ levels and the shift of the Fermi energy to a region with lower density of states, is responsible for the disappearance of magnetism. Figure \ref{Doping}(b) shows that the extra charge goes mainly to the As-Fe-As layer, as anticipated \cite {Takahashi-Nature}. 

\begin{figure}[h]
\includegraphics[width=66mm]{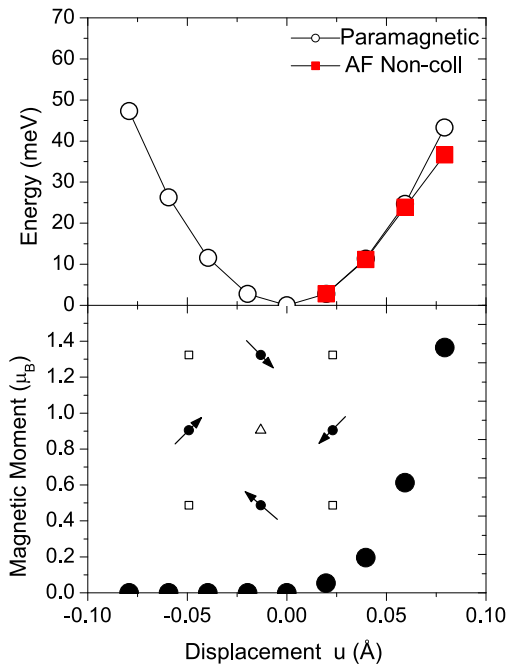}
\caption{(Color online) Variation of energy and magnetic moment with the amplitude $u$ of the $A_{1g}$-phonon at $\Gamma$ (vibration of As atoms perpendicularly to the Fe plane) for a doping of 0.275 electrons per formula unit. (a) Energy per formula unit in the paramagnetic and non-collinear magnetic configurations.  (b) Non-collinear magnetic moment of the Fe atoms. Their orientations are indicated in the inset, where squares and triangles stand for As atoms above and below the Fe plane. } \label{figure3}
\end{figure}

To address the electron-phonon interaction in this system, we have considered the symmetric out-of-plane FeAs $A_{1g}$ mode at $k_{||}=0$, in which the Fe atoms remain fixed whereas the As atoms move perpendicularly to the FeAs layers, expanding and compressing the Fe-As bonds. Figure \ref{figure3} shows the calculated total energy and magnetic moment, as a function of the expansion (positive $u$) or compression (negative $u$) of the Fe-As interplanar distance, for an extra concentration of 0.275 electrons per formula unit. Several points are worth mentioning:
\textit{i)} Expansion of the Fe-As distance induces magnetism in the Fe atoms.
\textit{ii)} In this case the arrangement of the magnetic moments at the Fe atoms is non-collinear, such that the As-mediated interaction remains antiferromagnetic, while there is no longer magnetic frustration for the direct Fe-Fe interaction, as in the AF1 order.
\textit{iii)} Compression of the Fe-As planes does not induce magnetism, and this asymmetry makes the vibration strongly anharmonic. 
The energy per formula unit can be fitted by the fourth order polynomial $E(u) = \hbar \omega (x^2 - 0.107 x^3 - 0.064 x^4)$, where $\hbar \omega = 29.0$ meV is the calculated phonon energy, $x \equiv u/u_0$ and $u_0 = 0.062$ \AA.
Thus, the restoring force is +0.91 and -0.68 eV/\AA\ for $u=$-0.06 and +0.06 \AA, respectively. 
\textit{iv)} The magnetic moments appear even for small phonon amplitudes, such that $E(u) \leq \hbar \omega$ ($u \leq u_0$).
\textit{v)} For positive $u$, the non collinear solution is less than $\sim 1$ meV lower than the AF1 configuration, which has very similar magnetic moments.
\textit{vi)} A similar, although weaker, effect has been obtained in the calculation of the asymmetric $A_{2u}$ mode.

\begin{figure}
\includegraphics*[width=78mm]{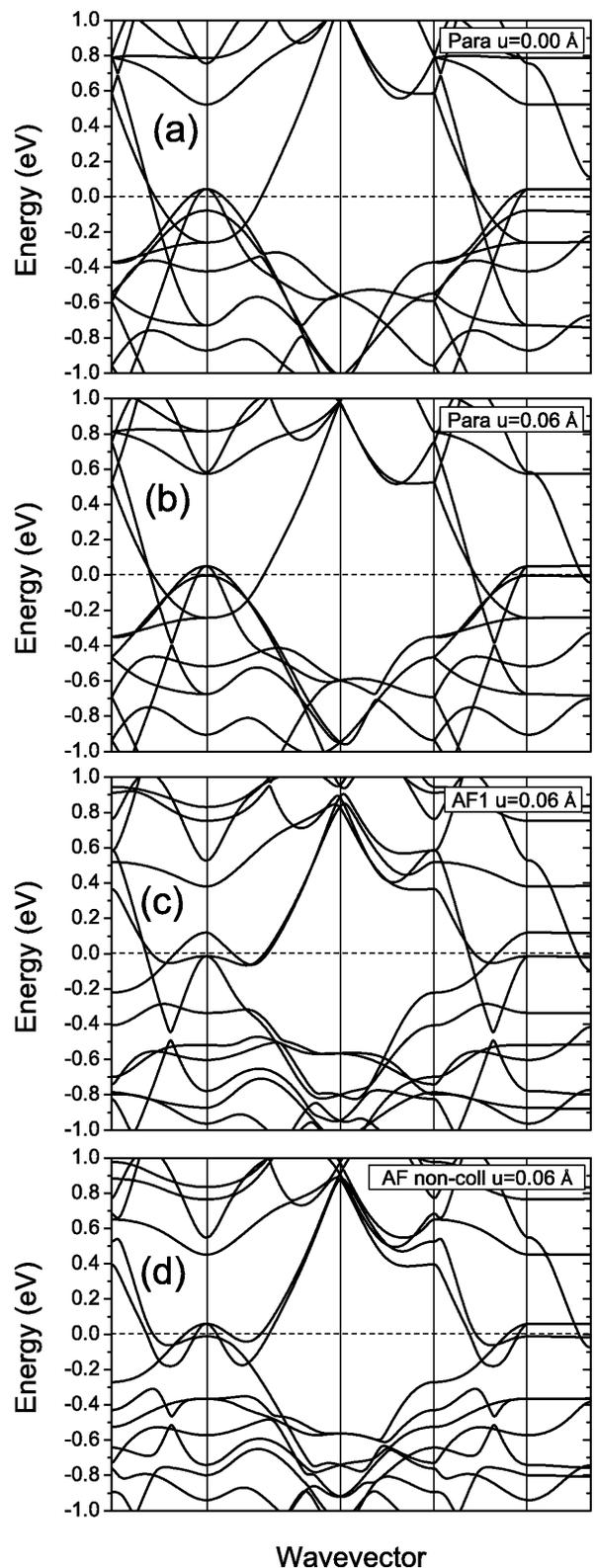}
\caption{Electronic band structure for LaOFeAs, doped with 0.275 electrons per formula unit, for different magnetic configurations (Paramagnetic, AF1, and AF non collinear, see text) and vibration displacements $u$ of the $A_{1g}$ phonon at $\Gamma$. The Brillouin zone is for the $(\sqrt{2}a\times\sqrt{2}a \times {c}) $ supercell. The zero of energy is at the Fermi level.} \label{Bands}
\end{figure}

\begin{figure}
\includegraphics*[width=84mm]{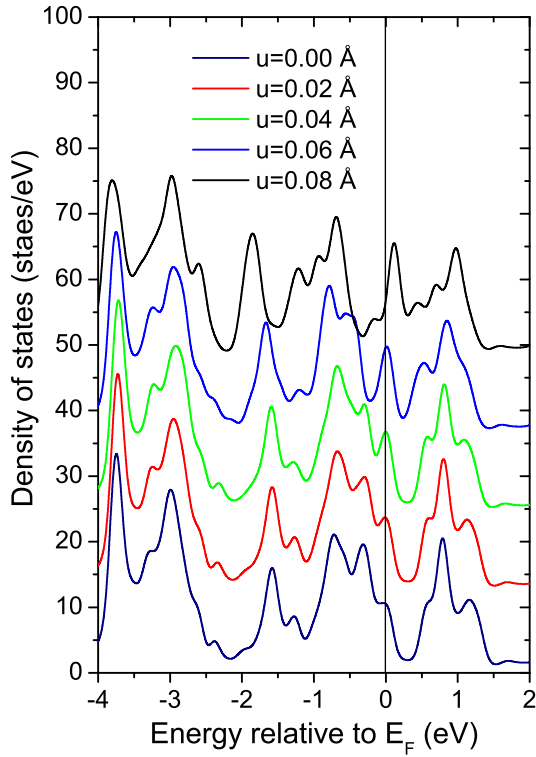}
\caption{(Color online) Densities of states of the stable non-collinear antiferromagnetic sate, for different vibration displacements $u$ of the $A_{1g}$-phonon at $\Gamma$. For clarity, the curves have been shifted and smoothed with a gaussian broadening of 0.1 eV. Lower to upper curves correspond to increasing values of $u$.} 
\label{DOS}
\end{figure}

Next we study how the $A_{1g}$ vibration affects the electronic structure in detail. Figure \ref{Bands} shows the calculated band structure, with an electron doping of 0.275, for different magnetic states and vibration displacements. The electron and hole pockets appear both at $\Gamma$, since the calculations were performed with a $(\sqrt{2}a\times\sqrt{2}a \times {c})$ supercell, with the primitive M point folded to $\Gamma$. The band structure is not much perturbed in the paramagnetic state (Figure \ref{Bands}(b)): the maximum deformation potential is $\sim 2$ eV/\AA\ and it occurs for states far from the Fermi level $E_F$. Around $E_F$ this value is much smaller, in agreement with Boeri et al \cite{Boeri}. In contrast, the electronic structure is strongly perturbed in both the AF1 and non-collinear antiferromagnetic states, which remove several band degeneracies (Figures \ref{Bands}(c)and (d)). In addition, antiferromagnetism removes states from about 0.3 eV below $E_F$, while a new peak in the density of states appears close to $E_F$ (Figure \ref{DOS}). This new peak is due to to the flattening of the electron pocket bands at the $\Gamma $ point, right at $E_F$.
The peak size increases with $u$, and it crosses $E_F$ for large $u$ (see Figure \ref{DOS}). The analysis of the partial density of states reveals that the states associated with this peak are fully spin polarized and entirely localized in the Fe atoms.

Finally, we suggest a possible general mechanism for BCS-like high-temperature superconductivity in antiferromagnetic compounds. 
It is based on driving them very close to a critical point or line of the antiferro-to-paramagnetic transition, in the pressure-doping phase diagram (we assume here $T=0$).
Under these conditions, the lattice distorsion of some phonons will switch antiferromagnetism on and off, thus enhancing dramatically their deformation potential and the $\lambda$ factor \cite{yildirim-2}.
In practice, chemical doping is inhomogeneous in the nanoscale, what implies that the superconducting band gap will vary spatially, or even that patches of normal and superconducting regions will coexist \cite{pan}.

In summary, we have shown that the electron-phonon interaction in paramagnetic electron-doped LaOFeAs is more complex and, probably, dramatically larger than previously predicted. The complexity arises from the vicinity of the system to magnetism and, in particular, with the abrupt appearance of phonon-induced magnetic moments at the Fe atoms, in an antiferromagnetic As-mediated configuration. The asymmetry of magnetism, with respect to the compression or expansion of the Fe-As bonds, makes the vibrations anharmonic. These results suggest that electron-phonon coupling must be carefully revised, before ruling out its connection with high-temperature superconductivity, in layered compounds close to an antiferromagnetic transition. 
Calculations of the complex electron-phonon interaction are under way, in order to address quantitatively this proposed mechanism.

We are indebted to  J. V. Alvarez, E. Anglada, E. Artacho G. Gomez-Santos, H. Suderow and S. Vieira for helpful discussions. This work was supported by the Spanish Ministry of Science and Innovation through grants FIS2006-12117 and CSD2007-00050.

\bibliographystyle{apsrev}
\bibliography{FeAsLaO}

\end{document}